\documentclass[aps,prb,twocolumn,showpacs,groupedaddress]{revtex4}
\usepackage{graphicx}

\begin{document}

\title{Floquet scattering theory of negative magneto-conductance of 2DEG under microwave radiation   }

\author{Kang-Hun Ahn}
\affiliation{ Department of Physics, Chungnam National University,
 Daejeon 305-764, Republic of Korea}
\date{ \today}

\begin{abstract}
We develop a theory of magneto-transport properties of
two-dimensional electron gas in cylindrical geometry where finite
bias and microwave are locally applied. By invesitgating the
Landauer conductance based on Floquet theory, we obtain the
conductance without perturbation analysis. We show that the dc
conductance becomes negative when dc electric field is applied
inside the 2DEG with sufficient microwave power. A positive sign
of the current is restored by increasing dc bias. We analyze the
radiation induced non-equilibrium distribution function with
respect to the frequency of the applied microwave.
\end{abstract}
 \maketitle

\section{Introduction}

Recently, radiation-induced vanishing magneto-resistance observed
in high-mobility 2DEG has attracted a great deal of
attraction\cite{klitzing,zudov,yang}. One of the current
understanding on the phenomena is based on the possible negative
resistance in certain conditions which evolves into certain
``zero-resistance(conductance)" state\cite{shi,andreev}. The idea
of negative resistance(conductance) was originally invoked long
time ago by Ryzhii\cite{ryzhii,ryzhii2}. Recent work of Durst et
al. proposed the negative resistance in terms of the photon
assisted electron-impurity scattering\cite{durst}. Subsequent
theoretical works explained the negative resistance in terms of
the radiation induced change of electron
distribution\cite{dmitriev}.

 An alternative way of explanation of the "zero resistance" exists.
Based on a semiclassical approach using a phenomenological damping
parameter, I\~narrea et al. directly proposed vanishingly small
positive resistance\cite{inarrea}. According to
Ref.\cite{inarrea}, the physical origin of the zero resistance is
that the electric current is blocked by Pauli exclusion principle
when the ``Fermi level" oscillates sufficiently.

 So far, most of theoretical study has been based on the
perturbation analysis. The purpose of this paper is to study the
radiation-induced electron transport without relying on the
perturbation analysis. We will not use 'Fermi golden rule' for the
transport which sometimes gives ambiguous conclusion on the
validity of the theory.

By this means, we study the magneto transport of electrons in
Landauer geometry. The effects of Landau level broadening are
realized via finite guiding center potential and the finite bias
gives rise to the shift of the density of states. We do not assume
 equilibrium function for a localized states defined by magnetic
 fields. Because when disorder in 2DEG play a role in magneto
 transport, (quasi-)extended states opens transport
 channels\cite{chalker}
 therefore the occupation of a localized state do not follow
 Fermi-Dirac distribution function.
We found the negative conductance in consistent with the previous
calculations showing negative resistance
\cite{ryzhii,ryzhii2,durst,dmitriev,shi}. Our model study
indicates that

(i) the negative conductance indeed exists even in purely quantum
mechanical calculations without perturbation analysis.

(ii) If the magnetic orbitals open transport channels via disorder
scattering, the Pauli exclusion principle do not ensure the zero
conductance.

 We also investigated the nonlinear effects and the breakdown of
the negative conductance as a function of DC bias. We analyze the
radiation induced non-equilibrium distribution functions for
temperatures lower than Landau level spacing. This paper is
organized as following. In section II, we introduce the model and
related formalism for the calculations of transport properties. In
section III, we show the linear negative conductance indeed exists
in certain conditions. It will be shown that, however, if the dc
bias do not change the density of states, there is no negative
conductance. The nonlinear properties will be presented in section
IV, and the radiation induced non-equilibrium distribution
function will be presented in section V.

\section{General Formalism}

\subsection{The Model}

 We consider a 2DEG confined in a cylinder with radius R subjected to a transverse magnetic field
$  B$. (See Fig. 1.) By doing this we study magneto-transport in
periodic boundary condition as in Corbino disc
geometry\cite{yang}. Similar model for 2DEG with unidirectional
modulation has been studied in Ref.\cite{dietel}. We study
Landauer conductance in the system where the static electric field
$E$ and oscillating field are applied only restricted regime as
shown in Fig. 1. In usual 2DEG subjected to a transverse magnetic
field, the longitudinal transport does not arise without disorder
scattering,  because in that case the electron merely performs a
localized cyclotron motion. To mimic the long-range smooth
disorder potential which exits in high-mobility
samples\cite{klitzing,zudov}, we introduce a scattering potential
$V=-V_{0}\cos\phi$ (Here we use cylindrical coordinates
$(\rho,\phi,z)$). By doing this we could define transport channels
in the 2DEG system.

\begin{figure}
\includegraphics[width=7cm,height=5.5cm]{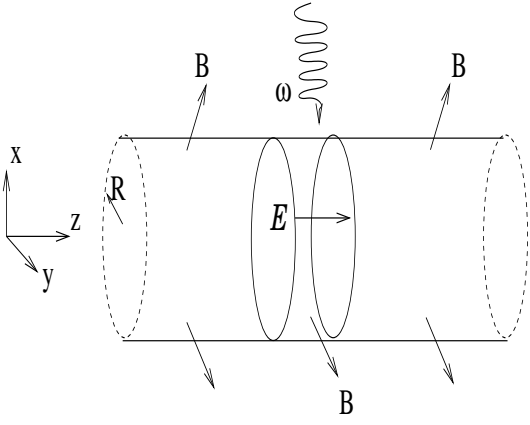}
 \caption{
A schematic figure for the model considered in this work.
 }
\end{figure}

In the absence of applied static electric field and microwave, the
Hamiltonian reads
\begin{eqnarray}
\label{hamilt}
H=\frac{1}{2m^{*}}\big({\bf p}-e {\bf A}  \big)^{2}-V_{0}\cos \phi,
\end{eqnarray}
were $m^{*}$ is the band effective mass of the electron in the
host materia and  the magnetic gauge in use is ${\bf A}=B z
\hat{\phi}$. For convenience, from here on we assign  $e=|e|$ and
$\hbar=1$. The eigen wavefunctions for  $V_{0}=0$ are $
\psi_{nm}(z,\phi) =\frac{1}{\sqrt{2\pi}}e^{i
m\phi}\phi_{n}(z-z_{m} )$ where $\phi_{n}$ is n th eiegen
wavefunction of harmonic oscillator with frequency
$\omega_{c}=eB/m^{*}$ and $z_{m}=m l_{B}^{2}/R$, and
$l_{B}=\sqrt{1/eB}$ is the magnetic length.

We neglect  the Landau level mixing caused by guiding center potential
$V(\phi)$ and DC electric field.
Then we get a tight binding model for the Hamiltonian
\begin{eqnarray}
 \nonumber
 H_{0}&=&\sum_{n,m} \big[n\omega_{c}+\delta_{m}\big]
c^{\dagger}_{n,m}c_{n,m}
\\
 &-&\frac{V_{0}}{2}\left(c^{\dagger}_{n,m+1}c_{n,m} + c^{\dagger}_{n,m-1}c_{n,m}\right),
\end{eqnarray}
where $c^{\dagger}_{nm}$ ( $c_{nm}$ ) is creation (annihilation)
operator for $\psi_{nm}(z,\phi)$. Here the DC electric field is
applied in the regime of $0 < z < Ml_{B}^{2}/R$ and gives finite
voltage bias $V$ which is dictated by
\begin{eqnarray}
\nonumber
\delta_{m}&=&0~~(m \le 0) \\
\nonumber
          &=&-eVm/M~~(0 < m < M) \\
\nonumber
          &=&-eV~~(  M \le m)
\end{eqnarray}
The transport channels are defined as eigenstates of the above
Hamiltonian for $V=0$. The eigenstates for zero bias, $V=0$, are
\begin{eqnarray}
|\psi_{n q}>=\sum_{m}\exp(i q m) c^{\dagger}_{nm} |0>,
\label{tchannel}
\end{eqnarray}
and the eigen energies are
\begin{eqnarray}
\epsilon_{n}(q)=n\omega_{c}-V_{0}\cos q.
\end{eqnarray}

Note that the DC electric field here modifies the transmission
probability and the density of states (DOS) as well. As we shall
see, it is crucial to consider the modification of DOS for the
negative conductance. The conductance is positive definite without
the change of DOS by the electric field.

We consider AC electric field $E_{\omega}\sin\omega t \hat{k}$
along z-direction localized in
 $0 < z< Ml_{B}^{2}/R$.
By replacing ${\bf A}$ with a time-periodic vector potential ${\bf
A}+\hat{k}\frac{E_{\omega}}{\omega}\cos\omega t$ in
Eq.(\ref{hamilt}), we obtain the time-dependent Hamiltonian
$H(t)=H_{0}+h(t)$ where
\begin{eqnarray}
\nonumber
 h(t)&=&\sum_{n=0}^{\infty}\sum_{m=1}^{M-1} i
\frac{eE_{\omega}\cos\omega t}{\sqrt{2}m^{*}\omega l_{B}}
 \sqrt{n}
\left(c_{n-1,m}^{\dagger}c_{n,m}-c^{\dagger}_{n,m}c_{n-1,m} \right)
\\&+& \left( \frac{eE_{\omega}\cos \omega t}{\sqrt{2 m^{*}}\omega}        \right)^{2}
 c_{n,m}^{\dagger}c_{n,m}.
\end{eqnarray}

\subsection{Floquet scattering theory}

Relying on the Floquet theorem, we write the eigenstates of the
time-periodic Hamiltonian $H(t)$ in terms of the coefficients
$\phi_{nm}^{(l)}$ ;
\begin{eqnarray}
\nonumber
\big| \psi(t) \big> &=& e^{-iEt}\big|\phi(t)\big>= e^{-iEt}\sum_{l}e^{-il\omega t}\big|\phi_{l}\big>\\
& =& \sum_{l}e^{-i(E+l\omega) t}\sum_{nm}\phi_{nm}^{(l)}
c_{nm}^{\dagger}\big|0\big> \label{wavefunc}
\end{eqnarray}
Then the time-dependent Schr\"odinger equation
$\frac{\partial}{\partial
t}\big|\psi(t)\big>=H(t)\big|\psi(t)\big>$ reads
\begin{eqnarray}
\nonumber (E+l \omega)\phi^{(l)}_{nm}
&=&\sum_{l^{\prime}n^{\prime}m^{\prime}}\frac{2\pi}{\omega}\int_{0}^{\frac{2\pi}{\omega}}dt
 e^{i(l-l^{\prime})\omega t}\times \\ && <nm|H(t)|n^{\prime}m^{\prime}>
\phi^{(l^{\prime})}_{n^{\prime}m^{\prime}} \label{floquet-eq}
\end{eqnarray}

\begin{figure}
\includegraphics[width=7cm,height=5.5cm]{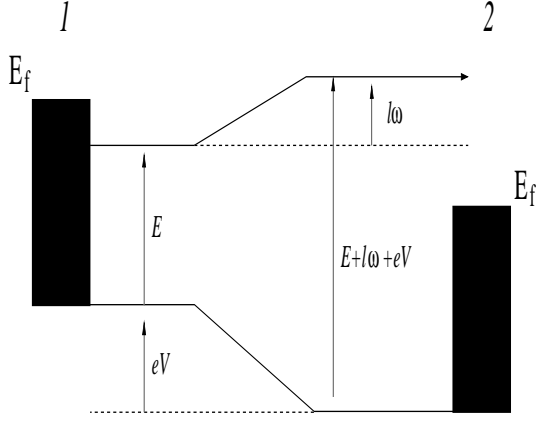}
 \caption{
A schematic figure for the energy states considered in this work.
 }
\end{figure}

Consider an incoming electron in the transport channel of energy
$E+L \omega $ in the N th Landau level in Eq.(\ref{tchannel}). For
$m \le 0$, the eigenstate is the sum of the incoming wave
$\big|\phi_{L,q_{LN}}\big>$
 and the reflected wave
$\big|\psi_{l, -q_{ln}}\big>$ with probability amplitude
$r_{0N}^{ln}$
\begin{eqnarray}
\big| \psi(t) \big> &=& e^{-iE_{L}t}\big|\phi_{L,q_{LN}}\big>+
\sum_{ln}r_{LN}^{ln}e^{-iE_{l}t}\big|\psi_{l, -q_{ln}}\big>,
\end{eqnarray}
where
\begin{eqnarray}
E_{l}=E+l\omega =n\omega_{c}-V_{0}\cos q_{ln}.
\end{eqnarray}
Thus, the coefficients  $\phi_{nm}^{(l)}$ for $m \le 0$ are
written as
\begin{eqnarray}
\phi_{nm}^{(l)}=\exp(iq_{LN}m)\delta_{lL}\delta_{nN}+r_{LN}^{ln}e^{-iq_{ln}m}.
\label{r-eq}
\end{eqnarray}

In the right-hand side of conductor ($ m \ge M$), the eigenstate
$|\psi(t)>$ and its corresponding coefficients $\phi_{nm}^{(l)}$
are written as
\begin{eqnarray}
\big|\psi(t)\big> &=&
\sum_{ln}t_{LN}^{ln}e^{-i(E+l\omega)t}\big|\psi_{n, k_{ln}}\big> \\
\phi_{nm}^{(l)}&=&t_{LN}^{ln}e^{ik_{ln}m},
\label{t-eq}
\end{eqnarray}
where
\begin{eqnarray}
E_{l} =n\omega_{c}-V_{0}\cos k_{ln}-eV.
\end{eqnarray}

By inserting Eqs.(\ref{r-eq}) and (\ref{t-eq}) into
Eq.(\ref{floquet-eq}), we obtain equations for the coefficients of
the transmission $t_{LN}^{ln}$ and reflection $r_{LN}^{ln}$.
Detail of the calculation can be found in the Appendix. The
transmission function from the energy $E$ in lead 1 to the the
energy $E+l\omega+eV$ in lead 2 is given by the sum of the
transmission probability over all possible spatial channels in the
leads 1 and 2 (See Fig. 2);
\begin{eqnarray}
\bar{T}_{21}(E+l\omega+eV,E+L\omega)
= \sum_{n,N}\big|t_{LN}^{ln}\big|^{2}\frac{\sin k_{ln}}{\sin q_{LN}}\Big|_{k_{ln}, q_{L,N}~real}
\end{eqnarray}

In the above equation, the sine  functions are appropriate
normalization factors for the the electric current,
 $J=\frac{e}{2}iV_{0}\sum_{m}
c^{\dagger}_{n,m+1}c_{n,m}-c^{\dagger}_{n, m-1}c_{n,m}$.

Assuming the both leads are in equilibrium given by the
Fermi-Dirac distribution $f_{FD}(E)=1/(\exp(\beta(E-E_{F}))+1)$,
we arrive at the current formula
\begin{eqnarray}
\nonumber I&=&\frac{2e}{h}\int_{-V_{0}}^{\infty} dE
\sum_{l=-\infty}^{\infty} \Big[ \bar{\rm T}_{21} \big( E+l\omega +
eV, E\big)-\\ && \bar{\rm T}_{12}\big(E+l\omega-eV,E\big)\Big]
f_{FD}(E)
\end{eqnarray}

Time-reversal symmetry in the Hamiltonian give us useful
information on the transmission function;
\begin{eqnarray}
\bar{\rm T}_{21}\big(  E+l\omega+eV,E) = \bar{\rm T}_{12}\big( E,
E+l\omega+eV ),
\end{eqnarray}
which leads to a simple form of the electric current;
\begin{eqnarray}
\nonumber
 I&=&\frac{2e}{h}\int_{-V_{0}}^{\infty}dE\sum_{l}
\bar{\rm T}_{21}\big( E+l\omega+eV,E\big)\\ &\times&
\big[f_{FD}(E)-f_{FD}(E+l\omega+eV) \big]
\label{c-formula}
\end{eqnarray}

From the above formula, one may notice that the electric current
of negative sign is attributed to inelastic transition processes
from occupied states of energy $E+l\omega+eV$ to unoccupied states
of energy $E$.

\section{Linear Conductance}

From the current formula in Eq.(\ref{c-formula}) we write the linear conductance
as
\begin{eqnarray}
\nonumber G &=&\frac{dI}{dV}\Big|_{V=0} = G_{ph}+G_{0} \\
\nonumber
 &=& \frac{2e}{h}\int dE
\sum_{l} \left[ \frac{\partial}{\partial V} \bar{\rm T}_{21}\big(
E+l\omega+eV,E\big) \right]_{V=0}
\\ &\times&
\big( f_{FD}(E)-f_{FD}(E+l\omega)\big)\\
\nonumber
&+&
\frac{2e}{h}\int dE \sum_{l}
\bar{\rm T}_{21}\big( E+l\omega,E\big)(-f^{\prime}_{FD}(E+l\omega))
\end{eqnarray}
Let us call the first part  photo-conductance $G_{ph}$ and the
second part $G_{0}$. At absolute zero temperature, $G_{0}$ is
written as simple Landauer formula;
\begin{eqnarray}
\nonumber G_{0}&=& \frac{2e^{2}}{h}\sum_{l} \bar{\rm T}_{21}\big(
E_{F},E_{F}+l\omega\big)
\\
&=& \frac{2e^{2}}{h}\sum_{l}\bar{\rm T}_{21}\big(
E_{F}+l\omega,E_{F}\big) \label{g0}
\end{eqnarray}

Note that $G_{0}$ is the conductance obtained when the applied
voltage drop is located only at the contacts between 2DEG and
leads. In this case there is no applied electric field in 2DEG. If
there were no electric field inside the conductor, the net current
$I$ could be written as
\begin{eqnarray}
\nonumber I&=&\frac{2e}{h}\int dE \sum_{l} \bar{\rm T}_{21}\big(
E+l\omega,E\big)\\&\times&
\left(f_{FD}(E-eV)-f_{FD}(E)\right)~~({\rm no~ electric
~field~in~2DEG}).
\end{eqnarray}
Then the conductance is given as in Eq.(\ref{g0}) for zero
temperature. Although $G_{0}$ is close to the conductance in the
case without microwave, they are not necessarily same. It is
instructive to note that, since the total transmission probability
$\sum_{l}\bar{\rm T}_{21}\big( E_{F}+l\omega,E_{F}\big)$ is always
positive definite, {\it the conductance can not be negative
without dc electric field inside} 2DEG.

\begin{figure}
\includegraphics[width=7cm,height=5.5cm]{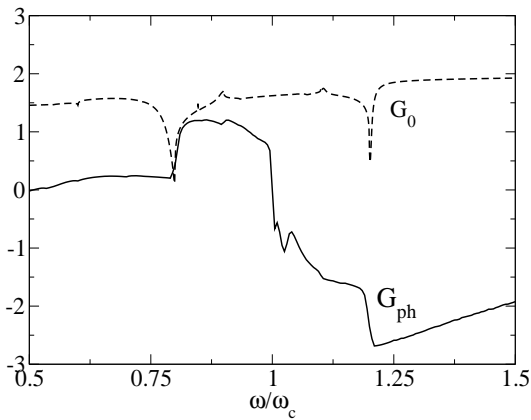}
 \caption{
The two contributions to conductance $G_{0}$ and $G_{ph}$ as
functions of the microwave frequency in unit of $2e^{2}/h$. The
chosen parameters are $V_{0}=0.7\omega_{c}$, $E_{\omega}=0.1
\times 2\sqrt{2}\omega_{c}/el_{B}$, and $E_{F}=2.5\omega_{c}$. The
Landau levels of $0 \le n \le 6$ were used for the calculations. }
\label{gomega}
\end{figure}

In Fig. \ref{gomega}, we show the numerical results of the
conductances as functions of the microwave frequency. The detail
of the calculation on the transmission probability was introduced
in the Appendix. Since we found that the all the features
discussed in this work are not significantly influenced by the
size of $M$, from here on, we focus on the case when the single
orbital is irradiated, $M=2$. The range of $l$'s for all the data
were chosen after convergence tests. ( For instance, $|l|\le 6$
was enough to produce Fig.\ref{gomega})

 The dips in
$G_{0}$ are due to so called "threshold anomaly". The threshold
anomaly arises whenever new transport channel is opening or
clsoing\cite{newton, reichl}. In our model, the density of states
$D(E)$ and the number of mode $M(E)$ of the 2DEG are given by
\begin{eqnarray}
D(E) &\propto&
\sum_{n=0}^{\infty}\frac{\Theta(E-n\omega_{c}+V_{0})\Theta(n\omega_{c}+V_{0}-E)}
{\sqrt{V_{0}^{2}-(E-n\omega_{c})^{2}}} \\
M(E)
&=&\sum_{n=0}^{\infty}\Theta(E-n\omega_{c}+V_{0})\Theta(n\omega_{c}+V_{0}-E)
\end{eqnarray}
For $0.5\omega_{c}< V_{0}<\omega_{c}$, $M(E)$ has maximum value of
2 when $E$ is located  in $(n+1)\omega_{c}-V_{0} <  E <n\omega_{c}
+ V_{0}$ where $n$ is a non-negative integer.Otherwise $M(E)$
equals to 1.
 In Fig. \ref{gomega} where $V_{0}=0.7\omega_{c}$, the dip of
$G_{0}$ at $\omega=0.8\omega_{c}$($1.2\omega_{c})$ is due to the
opening (closing)  of inter-Landau-level (intra-Landau-level)
transition where $M(E_{F}\pm \omega)$ increases (decreases) by 1.

The absolute negative conductance $G=G_{0}+G_{ph} $is clearly
shown in Fig.\ref{gomega} where the sudden sign reversal of
$G_{ph}$ at $\omega=\omega_{c}$ is responsible for the sudden
decrease of the net conductance.  The associated photo-current
$I_{ph}$ ($G_{ph}=dI_{ph}/dV$) reads
\begin{eqnarray}
\nonumber I_{ph}&=&\frac{2e}{h}\int dE \sum_{l} \bar{\rm
T}_{21}\big( E+l\omega+eV,E\big)
\\ &\times&
\big( f_{FD}(E)-f_{FD}(E+l\omega)\big). \label{iph}
\end{eqnarray}

 At zero temperature, the processes where both of
$E$,$E+l\omega+eV$ are below or above the Fermi level do not
contribute to the photo-current. This fact simplify our
consideration, because {\it when $eV <\omega$ we only have to
consider photon absorption}. Furthermore, the sign of
photo-current due to the process of given $l$ is given by the sign
of $l$.
 Now let us consider the most significant transition
process which arise when the energies of incoming electron and
outgoing electrons are all located in the regime where $M(E)$ is
maximum. Since $M(E)=M(E+\omega_{c})$, such an process is expected
when $(E+l\omega+eV)-E=l\omega+eV$ is close to integer multiple of
$\omega_{c}$. Therefore when $\omega \stackrel{<}{\approx}
\omega_{c}$, the term with $l=+1$ becomes most significant for
$eV>0$, which yields positive sign of the photo-current. On the
other hand, for $\omega \stackrel{>}{\approx} \omega_{c}$, the
condition
\begin{eqnarray}
l\omega+eV \approx \omega_{c}
\end{eqnarray}
is met when $l=-1$, which yields negative sign of photo-current.

\section{Nonlinear Properties}

\begin{figure}
\includegraphics[width=7cm,height=5.5cm]{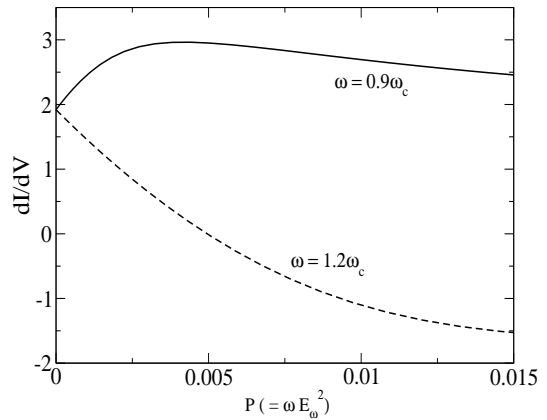}
 \caption{
The microwave power $P$  dependence of the differential
conductance $G=dI/dV$ at $V=0.01\omega_{c}/e$. $V_{0}=0.7$. $P$ is
in unit of $(2\omega_{c})^{3}/(el_{B})^{2}$}. \label{gpower}
\end{figure}

In Fig. \ref{gpower}, we plot the differential conductance $dI/dV$
as a function of microwave power
 $P=\omega E_{\omega}^{2}$.
For $\omega < \omega_{c}$ the conductance is enhanced by the
microwave  and eventually reach a maximum value and decreases. The
increase of the conductance is due to the strong absorption of the
single photon and the appearance of the nonlinear two-photon
process reduce the conductance as the power increases. For $\omega
> \omega_{c}$ the conductance decreases as the microwave power
increases and eventually reaches the negative conductance regime.
In contrast to the positive conductance showing non-monotonic
behavior, the negative conductance just show monotonic decreases.

\begin{figure}
\includegraphics[width=7cm,height=5.5cm]{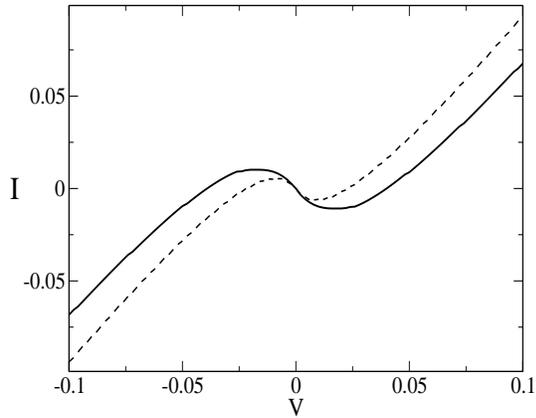}
 \caption{
The current-voltage characteristic curve. The solid line and
dashed line is for $\omega=1.22\omega_{c}$ and $1.3\omega_{c}$
respectively. The voltage is in units of
$\omega_{c}/e$.$E_{\omega}=0.1 \times 2\sqrt{2}\omega_{c}/el_{B}$
and $V_{0}=0.7$.
  }
\label{i-v}
\end{figure}

In Fig. \ref{i-v}, we plot the electric current as a function of
the applied bias voltage. The negative absolute conductance is
shown in a a region of finite bias and it is recovered as the bias
increases. The decrease of the photo current is the main reason of
the recovery of the negative conductance. According to previous
works by Ryzhii\cite{ryzhii3}, there is electric field $E_{0}$ at
which the conductivity becomes zero $\sigma(E_{0})=0$. The
electric-field breakdown arises at rather high electric field
$E_{0}l_{B}> {\rm max}(\omega-\omega_{c},\Gamma)$,where $\Gamma$
is the Landau level broadening. In our model, $\Gamma\sim V_{0}$.
As shown in Fig.\ref{i-v}, the electric-field breakdown appears at
rather small electric field where the breakdown field is roughly
an order of magnitude smaller.

\section{Radiation-induced non-equilibrium distribution function}

The electron distribution  becomes highly non-equilibrium due to
the electron scattering with the microwave and static electric
field. The distribution function of outcoming electron
$f_{out}^{(1,2)}$ in the lead 1 and 2 can be written as

\begin{eqnarray}
f^{(1)}_{out}(E)&=&\frac{1}{M(E)}\sum_{l}
\bar{T}_{12}(E,E_{l}+eV)f_{FD}(E_{l}+eV) \\ \nonumber &+&
\bar{T}_{11}(E,E_{l})f_{FD}(E_{l}) \\
f^{(2)}_{out}(E)&=&\frac{1}{M(E)}\sum_{l}
\bar{T}_{21}(E,E_{l}-eV)f_{FD}(E_{l}-eV) \\ \nonumber &+&
\bar{T}_{22}(E,E_{l})f_{FD}(E_{l})
\end{eqnarray}

\begin{figure}
\includegraphics[width=7cm,height=5.5cm]{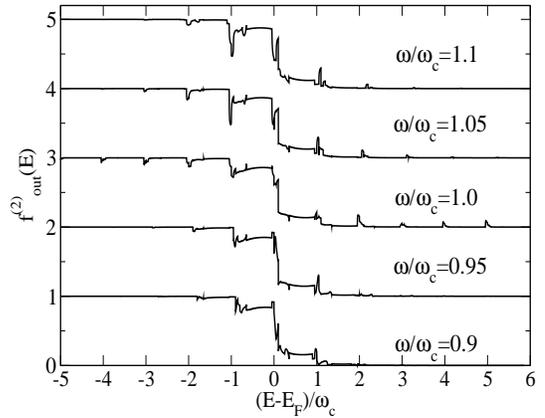}
 \caption{The non-equilibrium distribution function $f_{out}^{(2)}(E)$ for
$E_{\omega}=0.1 \times 2\sqrt{2}\omega_{c}/el_{B}$,
$V_{0}=0.7\omega_{c}$, $E_{F}=4.35\omega_{c}$, $eV=0.1\omega_{c}$.
} \label{fdist}
\end{figure}

In Fig.\ref{fdist}, we show an example of the distribution
function $f^{(2)}_{out}(E)$. It is shown that the distribution
function has signature of 'periodic' modulation by the microwave
irradiation only when $\omega$ is close to $\omega_{c}$.

In our Landauer approach where we assumed reflectionless contacts,
the non-equilibrium distribution function do not directly affect
the conductance. The non-equlibrium distribution function is a
result of the electron scattering and the scattered electron just
go away from conductors. For more realistic distribution function
where many scattering events are considered, the reader may refer
to Ref.\cite{dmitriev}. Our work might be useful for getting hint
on the zero temperature case which are not accessible within
semiclassical approximation.

\section{Conclusion}

We have developed the Floquet scattering theory of the negative
conductance of 2DEG in cylindrical geometry under microwave
radiation. Not only the quantum Hall system but also other
multi-channel conductors can have microwave induced negative
conductance. The necessary condition of the negative conductance
is that multi-channel conductors have dc electric field inside and
microwaves are irradiated. We demonstrated that the conductance is
always positive definite when the DC electric field do not affect
the density of states in conductors. The chemical potential
difference in contacts without modification of DOS by electric
field do not assure negative conductance even in the presence of
microwave.

The reason of the absence of the Pauli blocking\cite{inarrea} in
our work is that the magnetic orbitals are not merely localized
states but form transport channels due to disorder scattering.
Since the outgoing states and incoming states are different
states, those states do not have a relation of Pauli exclusion
principle. Situation will be different if the transport arises due
to hopping between localized states. In this case, Pauli exclusion
principle may play important role due to significant back
scattering events. Further theoretical works on the hopping
conduction with microwave are necessary to clarify the issue on
the Pauli blocking effect.

\section{Appendix}
In this appendix we develop the calculational method for the
Floquet scattering matrix in use in this work. We are going to
calculate the transmission(reflection) probability amplitude
$t_{0N}^{ln}$($r_{0N}^{ln}$) for the incoming electron with $l=0$
in $N$th Landau level to transmit(reflect) to the outcoming state
with $l$th channel in $n$th Landau level.

 To save the notations, we
introduce abstract vectors $\vec{\phi}_{m}$ which are defined as
the linear combinations the spinor $\vec{\chi_{nl}}$. The $(n,l)$
component of $\vec{\phi}_{m}$ denotes the coefficients
$\phi_{nm}^{(l)}$ for the wave-function in Eq.(\ref{wavefunc});
\begin{eqnarray}
\vec{\phi}_{m}=\sum_{n l}\phi_{nm}^{(l)}\vec{\chi}_{nl}.
\end{eqnarray}
Here $\vec{\chi}_{NL}$ means a spinor of which $(n, l)$ component
is 1 for $(n, l)=(N, L)$ and  zero for $(n,l)\neq (N, L)$.

We also introduce operators acting on the vectors;
\begin{eqnarray}
{\bf L_{0}}\vec{\chi}_{n l}&=&l\vec{\chi}_{n l} \\
{\bf L^{\pm}}\vec{\chi}_{n l}&=&\vec{\chi}_{n l\pm1} \\
{\bf H_{m}}\vec{\chi}_{nl}&=&(n\omega_{c}+\delta_{m})\vec{\chi}_{n
l}\\
{\bf N^{+}}\vec{\chi}_{nl}&=&\sqrt{n+1}\vec{\chi}_{n+1  l}\\
{\bf N^{-}}\vec{\chi}_{nl}&=&\sqrt{n}\vec{\chi}_{n-1  l}
\end{eqnarray}
The Schr\"odinger equation in Floquet basis in
Eq.(\ref{floquet-eq}) can be written as
\begin{eqnarray}
\nonumber (E+\omega{\bf L_{0}})\vec{\phi}_{m}&=& -\frac{V_{0}}{2}
(\vec{\phi}_{m-1}+\vec{\phi}_{m+1}) +{\bf H_{m}}\vec{\phi}_{m}\\
\nonumber &+&i\lambda({\bf N^{-}}- {\bf N^{+}})( {\bf L^{-}}+ {\bf
L^{+}} )\vec{\phi}_{m}\\
 &+&\lambda^{2}(  {\bf L^{-}}+ {\bf
L^{+}} )^{2}\vec{\phi}_{m},~~~(1\le m\le M-1)
\end{eqnarray}
where
$\lambda=\frac{eE_{\omega}l_{B}}{2\sqrt{2}\omega_{c}\omega}$. For
$m\le0$ and $m \ge M$, the above equation is simply modified by
setting $\lambda=0$.

Now we construct transfer matrix for
$(\vec{\phi}_{m},\vec{\phi}_{m-1})$. This task can be done similar
to the well-known transfer matrix calculation for Anderson
localization problem;
\begin{eqnarray}
\left(%
\begin{array}{c}
  \vec{\phi}_{m+1} \\
  \vec{\phi}_{m} \\
\end{array}%
\right)
=
\left(%
\begin{array}{cc}
  {\bf K_{m}}+{\bf F} & -{\bf 1} \\
  {\bf 1} & {\bf 0} \\
\end{array}%
\right)
\left(%
\begin{array}{c}
  \vec{\phi}_{m} \\
  \vec{\phi}_{m-1} \\
\end{array}%
\right)
\end{eqnarray}
where $m=1,2,...,M-1$ and
\begin{eqnarray}
{\bf K_{m}}&=&\frac{2}{V_{0}}\big[-(E+\omega{\bf L_{0}})+{\bf
H_{m}}
\big] \\
{\bf F}&=&\frac{2\omega_{c}}{V_{0}}\big[ i\lambda ({\bf N^{-}}-
{\bf N^{+}})( {\bf L^{-}}+ {\bf L^{+}} )+\lambda^{2}(  {\bf
L^{-}}+ {\bf L^{+}} )^{2} \big]
\end{eqnarray}

Successive multiplication of the transfer matrix give us the
following relation.
\begin{eqnarray}
\left(%
\begin{array}{c}
  \vec{\phi}_{M+1} \\
  \vec{\phi}_{M} \\
\end{array}%
 \right)
 &=&
 {\bf T}
\left(%
\begin{array}{c}
  \vec{\phi}_{0} \\
  \vec{\phi}_{-1} \\
\end{array}%
 \right),
 \label{transfer-wv}
 \end{eqnarray}
 where
 \begin{eqnarray}
 \nonumber
{\bf T}&=&
\left(%
\begin{array}{cc}
  {\bf K_{M}} & -{\bf 1} \\
  {\bf 1} & {\bf 0} \\
\end{array}%
\right)
 \prod_{m=1}^{M-1}
\left(%
\begin{array}{cc}
  {\bf K_{m}}+{\bf F} & -{\bf 1} \\
  {\bf 1} & {\bf 0} \\
\end{array}%
\right) \\ &\times&
\left(%
\begin{array}{cc}
  {\bf K_{0}} & -{\bf 1} \\
  {\bf 1} & {\bf 0} \\
\end{array}%
\right).
\end{eqnarray}

From Eqs.(\ref{r-eq}) and (\ref{t-eq}), the transmission and
reflection amplitudes are written as
\begin{eqnarray}
\vec{\chi}_{nl}\cdot\vec{\phi}_{M+1} &=&
t_{0N}^{ln}\exp(i(M+1)k_{ln}) \\
\vec{\chi}_{nl}\cdot\vec{\phi}_{M} &=&
t_{0N}^{ln}\exp(iM k_{ln}) \\
\vec{\chi}_{nl}\cdot\vec{\phi}_{0}
&=&\delta_{l0}\delta_{nN}+r_{0N}^{ln} \\
\vec{\chi}_{nl}\cdot\vec{\phi}_{-1}&=&
\exp(-iq_{0N})\delta_{l0}\delta_{nN}+r_{0N}^{ln}\exp(iq_{ln}).
\end{eqnarray}
By solving the above coupled linear equations with
Eq.(\ref{transfer-wv}), we obtain $t_{0N}^{ln}$ and $r_{0N}^{ln}$.

 \vspace{10cm}
\begin{acknowledgements}
This work was supported by Chungnam National University through
Internal Research Program. I thank Prof. P. Fulde for helpful
discussion and Max Planck Insititute for Physics of Complex
Systems where part of this work has been done.
 \end{acknowledgements} 

\end{document}